\documentstyle[12pt]{article}

\newcommand{\mysection}[1]{\setcounter{equation}{0}\section{#1}}

\newcommand{\beq}{\begin{equation}}
\newcommand{\eeq}{\end{equation}}
\newcommand{\ber}{\begin{eqnarray}}
\newcommand{\eer}{\end{eqnarray}}

\newcommand{\eeql}[1]{\label{#1}\eeq}%\marginpar{\vskip-1cm\tiny #1}}
\newcommand{\eerl}[1]{\label{#1}\eer}%\marginpar{\vskip-1cm\tiny #1}}

\newfont{\Bbb}{msbm10 scaled 1200}     %instead of eusb10
\newcommand{\mathbb}[1]{\mbox{\Bbb #1}}
\def\IR{{ R}}

\input epsf

\newcommand{\figin}[2]{
\begin{figure}[t]
\centerline{\hbox{\epsffile{#1.eps}}}
\centerline{\parbox{12cm}{\caption{#2\label{#1}}}}
\end{figure}}

\newcommand{\fig}[1]{Fig.~\ref{#1}.}
\newcommand{\figur}[2]{\fig{#1}\figin{#1}{#2}}

\newcommand{\vev}[1]{{\left< {#1} \right>}}

\newcommand{\Tr}{{\rm Tr\,}}

\newcommand{\cD}{{\cal D}}

\newcommand{\cN}{{\cal N}}
\newcommand{\cP}{{\cal P}}

\newcommand{\half}{{1\over 2}}

\newcommand{\preprint}[1]{\begin{table}[t]  %%
               \begin{flushright}               %%
               {#1}                             %%
               \end{flushright}                 %%
               \end{table}}                     %%
\renewcommand{\title}[1]{\vbox{\center\LARGE{#1}}\vspace{5mm}}
\renewcommand{\author}[1]{\vbox{\center#1}\vspace{5mm}}
\newcommand{\address}[1]{\vbox{\center\em#1}}
\newcommand{\email}[1]{\vbox{\center\tt#1}\vspace{5mm}}

\begin{document}

\begin{titlepage}
\preprint{hep-th/0010274 \\
USC/00-05\\
CITUSC/00-057\\
NSF-ITP-00-119}

\title{An Exact Prediction of ${\cal N}=4$ SUSYM Theory for String Theory}

\author{Nadav Drukker$^1$, David J. Gross$^2$}

\address{
$^1$CIT-USC Center for Theoretical Physics,
%Department of Physics and Astronomy,
University of Southern California,\\
Los Angeles, CA 90089-2535
\\
$^2$Institute for Theoretical Physics, University of California\\
Santa Barbara, CA 93106
}

\email{drukker@usc.edu, gross@itp.ucsb.edu}

\abstract{We propose that the expectation value of a
circular  BPS-Wilson loop in ${\cal N}=4$ SUSYM
can be calculated exactly, to all orders in a $1/N$ expansion and to
all orders in $g^2N$. Using the AdS/CFT
duality, this result  yields a
prediction of  the value of  the string amplitude with a circular
boundary to all orders in $\alpha'$ and to all orders
in $g_s$.  We then compare this result with string theory.
We find that the gauge theory  calculation, for  large $g^2N$ and to
{\it all} orders in the $1/N^2$
expansion does agree
with the leading   string theory  calculation, to all orders in $g_s$
and to lowest order in $\alpha'$. We also find a relation between
the expectation value of   any closed smooth Wilson loop  and  the loop
related to it by an inversion
that takes a point along the loop to infinity, and compare this result,
again successfully,
with string theory.

}
\end{titlepage}

%%%%%%%%%%%%%%%%%%%%%%
\mysection{Introduction}
%%%%%%%%%%%%%%%%%%%%%%

There have been many tests of the conjectured duality  of ${\cal
N}=4$ supersymmetric Yang-Mills theory (SUSYM)
     with type IIB string theory in $AdS_5\times S^5$ background.
However, since the duality relates
gauge theory with coupling $g^2$ and gauge group of rank $N$ to type
IIB string theory in
an $AdS$ background with radius $R^2=\sqrt{g^2N}\alpha'$ and string coupling
$4\pi g_s=g^2$, the only precise tests have
been of quantities so protected by supersymmetry that they receive no
perturbative or non-perturbative corrections.
It is easy to calculate quantities in the gauge theory for weak
coupling---but these yield predictions for string theory in
a very curved background, where there do not yet exist methods of
computation. Conversely, it is easy to
to calculate quantities in the string theory for weak coupling (large
$N$) and large curvature (or small $\alpha'$)
---but these yield predictions for the gauge theory for large $N$ and
large $g^2N$, for which there are no
reliable methods of computation. In neither case, so far, is there a
prediction on either side that holds for all $N$ and $g^2$.

We will suggest, in this paper, that the expectation value of a
circular  BPS-Wilson loop in ${\cal N}=4$ (SUSYM)
can be calculated exactly, to all orders in a $1/N$ expansion and to
all orders in $g^2N$. This then yields a
prediction of  the value of  the string amplitude with a circular
boundary to all orders in $\alpha'$ and to all orders
in $g_s$.  We then compare this result with string theory.
We find that the gauge theory  calculation, for  large $g^2N$ and to
{\it all} orders in the $1/N^2$
expansion does agree
with the leading   string theory  calculation, to all orders in $g_s$
and to lowest order in $\alpha'$.

Our result is an extension of
     a beautiful paper \cite{Erickson:2000af}, in which  Erickson,
Semenoff and Zarembo
calculated the contributions of rainbow graphs to the expectation value of
a circular Wilson loop in $\cN=4$ supersymmetric gauge theory. The result they
found was that:
\beq
\vev{W}_{\rm rainbow}=  {2\over\sqrt\lambda}I_1\left(\sqrt\lambda\right)\, ,
\eeql{ama}
where $\lambda=g^2N$ is the 't Hooft coupling and $I_1$ is a Bessel function.
For large $\lambda$ (\ref{ama}) behaves as
\beq
\vev{W}_{\rm rainbow}
\sim  \sqrt{2\over \pi} {e^{\sqrt\lambda}\over \lambda^{3/4}}\,\, .
\eeql{leading}
The expectation value of Wilson loops can also be calculated
using the Maldacena conjecture
\cite{Maldacena:1998re,Rey:1998ik,Maldacena:1998im}, and for the circular
Wilson loop one finds, to leading order in large $\lambda$, that
\cite{Berenstein:1999ij,Drukker:1999zq}
\beq
\vev{W}_{\rm circle}=e^{\sqrt\lambda} \ ,
\eeq
in agreement with  (\ref{leading}).

The authors of \cite{Erickson:2000af} conjectured that the rainbow
graphs gave the exact large
$N$ behavior of the circular Wilson loop and gave some evidence (a 2
loop calculation) to this effect.
We will outline a proof that the result, (\ref{ama}), is indeed exact to all
orders in $g^2N$ for $N=\infty$.
We will also generalize this result to all orders in the $1/N^2$ expansion.

How are we able to perform an exact calculation in strongly coupled
gauge theory? The reason turns out to be that
the circular Wilson loop is totally determined by an anomaly, a
conformal anomaly. As in other cases one
is able to calculate the anomaly exactly to all orders in the coupling.

To see this recall that the Wilson loop under discussion is the
     appropriate  supersymmetric Wilson loop
\beq
W={1\over N}\Tr\cP
\exp i\oint(A_\mu \dot x^\mu+i\Phi_i|\dot x| \theta^i)dt\,.
\eeql{wl}
where $A_\mu$ and $\Phi_i$ are the gauge fields and the scalars that couple
to $x^\mu(t)$, parameterizing the circle and to $\theta^i$ which is chosen to
be some constant unit vector in $\IR^6$.
This special Wilson loop is  locally supersymmetric.  If the contour
$ (x^{\mu}(t)$)
is a straight line   then the  Wilson line is globally a BPS object whose
expectation value is precisely one.  A straight line and a circle are
related by a conformal transformation. This fact was used by
\cite{Berenstein:1999ij} to find the minimal surface ending along a
circle.  If the expectation value of a Wilson loop  was
truly invariant under all conformal transformations then the
expectation value of a circular loop would also be one.
However, this is not the case. We will show that there are quantum
anomalies when one performs
the type of global conformal transformations necessary to   turn a
straight line into a circle.
These anomalies are responsible for the very non-trivial $g^2N$ and
$1/N$ behavior of the circular loop,
and as often is the case with anomalies, can be calculated exactly.

Accepting for the moment the result  of  (\ref{ama}) (for $N=\infty$),
we see  that acting on a straight line with a special
conformal transformation that changes it to a circle changes its
expectation value by a factor of:
${2\over\sqrt\lambda}I_1\left(\sqrt\lambda\right)$.  Since this
factor arises from an anomaly,
we will be able to argue  that this phenomenon is
much more general---the same happens for a general Wilson loop. That is
\beq
\vev{W}_{N=\infty}
={2\over\sqrt\lambda}I_1\left(\sqrt\lambda\right)\vev{  \tilde  W}\,,
\eeql{general}
where $W$ is any closed smooth Wilson loop  and $\tilde W$  is the loop
related to it by a special conformal transformation
that takes a point along $W$ to infinity. Even more, we will
generalize this result to all
orders in the $1/N^2$ expansion.

The fact that the expectation value of circular Wilson loops   and
straight line  Wilson loops (or
more generally  closed and  open loops related by conformal
transformations) are different should not be a surprise.
Large conformal transformations, such as an inversion
\beq
x^\mu\to {x^\mu\over x^2}\, ,
\eeq
are not  symmetries of $\IR^4$, since they exchange the point at infinity with
a point at a finite distance. They are a symmetry of the theory on $S^4$, which
     includes the point at infinity. On the sphere  there is no
distinction between a
circle and a line,  and the expectation value of either is the same as for a
circle on $\IR^4$.

There clearly could be a problem with the invariance under global
conformal transformations.
For example, a conformal transformation of a correlator of $n$ local operators
could take one of the points to infinity, and turn it into the correlator of
$n-1$ operators. Here we are seeing an analogous statement for Wilson
loops, by transforming the circle to the line, one point along the loop is
taken to infinity.
As such, one might guess that the difference between the line and the circle
is  the contribution   of the fields at a single point. In fact the authors of
\cite{Erickson:2000af} pointed out that (\ref{ama}) is equal to the
Wilson loop of the large $N$ Hermitean matrix model
\beq
{2\over\sqrt\lambda}I_1\left(\sqrt{\lambda}\right)
=\vev{{1\over N}\Tr \exp(M)}
={1\over Z}\int\cD M {1\over N}\Tr \exp(M)
\exp\left(-{2\over g^2}\Tr M^2\right)
\, ,
\eeq
and one   could associate the field $M$ with  the fluctuations  of
the fields at the point at
infinity.

We will demonstrate how equation (\ref{general}), and its finite $N$
generalization, can be proven in  Section 2.
The idea for the
proof is the following. Under a conformal transformation the gluon propagator
is modified by a total derivative. This is analogous to a gauge
transformation, and naively does not affect the gauge invariant loop.
However the gauge  transformation  is singular at the point that is
taken to infinity.
While the perturbative expansion is naively invariant under gauge
transformations,
we find that this invariance  breaks down at the singular point.  By
calculating the contribution from
the singularities we are able to show that it is given by a matrix
model. We did not complete the proof that the matrix model is quadratic,
but there are many indications that it indeed is. Under that assumption
we are able to evaluate the expectation value to all orders in
perturbation theory.
For large $N$ we will recover (\ref{general}), but our result yields
an exact relation for any $\lambda$
and any $N$. In the case of a circular loop we  derive an exact
expression for $\vev{W}$.

In Section 3 we  compare our results with the dual string theory. We find
that at the classical level the minimal area calculation shows the same
universal behavior under a conformal transformation.  In the case of
the circular loop,
where we are able to calculate in the gauge theory exactly, we argue
that order by order
in string perturbation theory, the leading contribution for small
$\alpha'$ - agrees with the gauge theory predictions. We also show 
That the agreement extends to large coupling where, after an S-duality 
transformation, it is given by a D1-brane.

In Section 4 we generalize the calculation to more general observables 
in the matrix model. Those correspond to Wilson loops wound multiple 
times around the circle.

The Appendices contain the details of the explicit evaluation of the
matrix model
that yields the precise form of our results.

%%%%%%%%%%%%%%%%%%%%%%%%%%%%%%%%%%%%%%
\mysection{The Gauge Theory Calculation}
%%%%%%%%%%%%%%%%%%%%%%%%%%%%%%%%%%%%%%

We shall explore the invariance of the Wilson loop under large
conformal transformations
order by order
in perturbation theory.

We expand the expectation value of the Wilson loop around some
contour $C$,  as defined in
(\ref{wl})
\ber
\vev{W_C}&=&\sum_{n=0}^\infty A_n \lambda^n\,, \nonumber\\
A_0
&=&1
\nonumber\\
      A_1
&=&\half\oint ds_1\oint ds_2{1\over N}\Tr
\left(-\dot x_1^\mu\dot x_2^\nu \vev{A_\mu(x_1)A_\nu(x_2)}
+|\dot x_1|\theta_1^i|\dot x_2|\theta_2^j \vev{\Phi_i(x_1)\Phi_j(x_2)}\right)
\nonumber\\
      A_2&=&\cdots
\eerl{ex}

We will work  in $\IR^4$, where the propagators  are
translationally invariant  and investigate the behavior  of
$\vev{W_C}$  under conformal
transformations that take the closed contour $C$ to $\tilde C$.  We will
compare $\vev{W_C}$  to
\beq
\vev{\tilde W_{\tilde C}}=\sum_{n=0}^\infty \tilde A_n \lambda^n\,.
\eeq
We could instead compare the gauge theory on $\IR^4$ to
the theory on $S^4$. In the latter we would use propagators that transform
covariantly under inversions. Those were studied in \cite{Adler:1972qq},
and are related to the Feynman gauge propagator by a singular gauge
transformation. The two computations turn out to be equivalent.

\subsection{Quadratic term}

Let us look at the first non trivial term in the expansion of the Wilson loop
and compare $\tilde A_1$ to $A_1$.
First consider the behavior of the propagators  under a
large conformal transformation.
In particular we shall examine the behavior under an   inversion about
the origin.
\beq
x^\mu\to {x^\mu\over x^2}= \tilde x^\mu\,.
\eeql{conftr}
All other large conformal transformations can be gotten by a
combination of an inversion
and small  conformal  transformations.
under inversion the scalar propagator
\beq
G_{ij}^{ab}(x_1,x_2)
=\vev{\Phi_i^a(x_1)\Phi_j^b(x_2)}
={g^2\over 4\pi^2}{\delta_{ij}\delta^{ab}\over(x_1-x_2)^2}\,,
\eeq
transforms to
\beq
\tilde G_{ij}^{ab}(\tilde x_1,\tilde x_2)
={g^2\over 4\pi^2}x_1^2 x_2^2{\delta_{ij}\delta^{ab}\over(x_1-x_2)^2}\,.
\eeq
Taking into account the fact that under  inversion $|\dot x|\to |\dot
x|/x^2$, the one scalar exchange contribution
\beq
|\dot x_1|\theta_1^i|\dot x_2|\theta_2^j \vev{\Phi_i(x_1)\Phi_j(x_2)}\,,
\eeq
to the  Wilson loop, is invariant under  inversion. However, if this
was the only term we would have to introduce
a ultraviolet cutoff to render the integral finite, and this could
spoil the conformal invariance.
     Indeed,
the Wilson loop in a non supersymmetric theory exhibits a perimeter law
in perturbation theory
\beq
W\sim g^2{L\over\epsilon}\,,
\eeq
which is definitely not invariant under conformal transformations. But the
inclusion of both the scalars and the gluons  in the Wilson loop exactly
cancels this
divergence \cite{Drukker:1999zq}.

The story with the gauge fields is more complicated, since under
inversion a vector field (of dimension one)   transforms as
\beq
\tilde V_\mu(\tilde x)=x^2 I_{\mu\nu}(x)V^\nu(x)\,,
\qquad
I_{\mu\nu}(x)=g_{\mu\nu}-2{x_\mu x_\nu\over x^2}\,.
\eeq
This can easily be derived by noting that
\beq
{\partial \over \partial \tilde x^\mu } =x^2I_{\mu\nu}(x){\partial
\over \partial  x^\nu },
\eeq
so that if $\Phi$ is a dimensionless scalar  field that transforms as
$\Phi(\tilde x)=\Phi(x)$,
then the dimension one vector field, $\partial_\mu \Phi(x)$, will
transform as above.

Thus the gluon propagator transforms  as
\beq
\vev{\tilde A_\mu^a(\tilde x_1) \tilde A_\nu^b(\tilde x_2)}
=x_1^2x_2^2 I_{\mu\rho}(x_1)I_{\nu\sigma}(x_2)
\vev{A_\rho^a(x_1) A_\sigma^b(x_2)}\,.
\eeq
We shall work, for convenience in Feynman gauge:
$\vev{A_\mu^a(x_1)A_\nu^b(x_2)}
={g^2\over4\pi^2}{g_{\mu\nu}\delta^{ab}\over(x_1-x_2)^2}$.
Then the transformed propagator is
\ber
\tilde G_{\mu\nu}^{ab}(\tilde x_1,\tilde x_2)
&=&{g^2\delta^{ab}\over 4\pi^2}{x_1^2x_2^2\over (x_1-x_2)^2}
\left(g_{\mu\nu}
-2{x_1^\mu x_1^\nu\over x_1^2}
-2{x_2^\mu x_2^\nu\over x_2^2}
+4{x_1\cdot x_2  x_1^\mu x_2^\nu\over x_1^2x_2^2}\right)
\nonumber\\
&=&{g^2\delta^{ab}\over 4\pi^2}x_1^2x_2^2
\Bigg({g_{\mu\nu}\over(x_1-x_2)^2}
+\half\partial_\mu^1\left(\ln (x_1-x_2)^2\partial_\nu^2\ln x_2^2\right)
\nonumber\\
&&\hskip.5in
+\half\partial_\nu^2\left(\ln (x_1-x_2)^2\partial_\mu^1\ln x_1^2\right)
-\half\partial_\mu^1\partial_\nu^2\left(\ln x_1^2\ln x_2^2\right)\Bigg)\,.
\eerl{transf}

Consequently, while we saw that the contribution of the scalars to
the Wilson loop was
invariant under inversion, the gluon
contribution, $\dot x_1^\mu\dot x_2^\nu
\vev{A_\mu^a(x_1)A_\nu^b(x_2)}$, is changed by  a total derivative:
\ber
&&\hskip-.3cm
{g^2\delta^{ab}\over 8\pi^2}\dot x_1^\mu \dot x_2^\nu
\Bigg[\partial_\mu^1\left(\ln (x_1-x_2)^2\partial_\nu^2\ln x_2^2\right)
+\partial_\nu^2\left(\ln (x_1-x_2)^2\partial_\mu^1\ln x_1^2\right)
-\partial_\mu^1\partial_\nu^2\left(\ln x_1^2\ln x_2^2\right)\Bigg]
\nonumber\\
&&={g^2\delta^{ab}\over 4\pi^2}\dot x_1^\mu \dot x_2^\nu\,
\partial_\mu^1\left(\ln {(x_1-x_2)^2\over |x_1|}\,
\partial_\nu^2\ln x_2^2\right)\, .
\eerl{total}

Since the modification of the gluon contribution
is  a total derivative, which is equivalent to
a gauge transformation, one might conclude that the inversion is a
symmetry of the Wilson loop.
     This would be the case, except that the gauge transformation   in
(\ref{total}) has potential singularities.
     We must  therefore
reexamine  the proof of gauge invariance of the perturbative expansion and see
whether it fails.

We are evaluating the integral
\beq
\tilde A_1-A_1
=-{1\over 16\pi^2}
\oint_C dx_1^\mu\oint_C dx_2^\nu\,
\partial_\mu^1\left(\ln {(x_1-x_2)^2\over |x_1|}\,
\partial_\nu^2\ln x_2^2\right)\,.
\eeql{int}
(Here we have  included the  contribution from the color indices
that gives a factor of
$
{1\over N}\Tr T^a T^a
={N\over 2}\,.
$)
There are two potential  singularities that we encounter when
doing the $x_1$ integral,  at $x_1=x_2$, and
     at $x_1=0$. The second singularity only occurs if the point
$x^\mu=0$ lies on the contour $C$.
To examine the behavior at the singularities we introduce a cutoff $\epsilon$.

First, consider the case where  $x^\mu=0$ lies on the contour $C$.
The contribution from $x_1=0$ is
\beq
-{1\over 16\pi^2}
\oint_C dx_2^\nu\,
\ln {(x_2+\epsilon)^2\over (x_2-\epsilon)^2}\,
\partial_\nu^2\ln x_2^2\,.
\eeq
Here $\epsilon$ is an infinitesimal vector tangent to the loop at the origin.
To perform the $x_2$ integral, we notice that for large $x_2$ the integrand
is of order $\epsilon$, the integrand can therefore be regarded as a delta
function concentrated at $x_2=\pm\epsilon$.
A similar term arises from regularizing the singularity at $x_1=x_2$, which is
also zero for $x_2$ far from the origin. So the only contribution comes from
the point $x_1=x_2=0$.

To find the contribution from the singular point one can use the expression
\beq
\int_\epsilon^\infty dx\,
{1\over x}\ln {x-\epsilon\over x+\epsilon}
=-{\pi^2\over 4}\,,
\eeq
to find that
\beq
\tilde A_1-A_1=-{1\over 8}\,.
\eeql{aone}

On the other hand if  $x^\mu=0$ does not lie  on the contour $C$  the
integral is not singular enough
and it vanishes. Thus in this case $\tilde A_1-A_1= 0$. Therefore we
conclude that
     under inversion  through the origin  the quadratic contribution to
the Wilson loop is invariant
if the original loop does not pass through the origin. Such an
inversion transforms
a closed contour into another closed contour.
On the other hand if $C$ passes through the origin the  transformed
Wilson loop
($\tilde C$) is now extended to infinity, it is a open
Wilson line that only meets at the point at infinity. In this case,
to quadratic order,
\beq
\vev{\tilde W_{\tilde C}}  -\vev{W_C} =-\lambda/8\ .
\eeq

A safer route to the same result is to evaluate  the  modification to
the propagator (\ref{transf}) directly, and
not use integration by parts. That way one does not encounter any
singularities. Let us do this for the two simplest examples. First we
look at a circle passing through the origin
\beq
x_1(s)=(1+\cos s,\ \sin s)\qquad x_2(t)=(1+\cos t,\ \sin t)\,.
\eeql{crc}
Under inversion this is mapped to the straight line: $x(s)={1\over
2}(1, \tan(s/2))$.
For this contour the modification of the gluon propagator contributes
to $\tilde W$ the amount
\ber
&& {\lambda\over 16\pi^2}\left[\dot x_1^\mu \dot x_2^\nu\,
\partial_\mu^1\left(\ln {(x_1-x_2)^2\over |x_1|}\,
\partial_\nu^2\ln x_2^2\right) +( {x_1 \to x_2})\right]= \nonumber \\
&&  {\lambda\over 16\pi^2}\left[ - \left({ 2\sin{t}\over
4\sin^2{t\over 2}}\right)
\left({2 \sin{s}\over 4\sin^2{s\over 2}}\right)
+ {2\sin(s-t)\over 4\sin^2({s-t\over 2})}\left({2 \sin{s}\over
4\sin^2{s\over 2}}\right)
\left({ \sin{t}\over 4\sin^2{t\over 2}} \right)
\right]=-{\lambda\over 32\pi^2} \,,\nonumber
\\
     \eer
which, when integrated over the circle,
gives the result of  (\ref{aone}).

It is even simpler to take a straight line that does not pass through the
origin
$
x(s)=(1,\ s).$
Under the inversion it is mapped to a circle, of radius $1/2$ whose
origin is at $(1/2,0)$,
and the point at infinity is
mapped to the origin. Therefore we expect a contribution from the point at
infinity that is exactly opposite to the previous calculation. Indeed
\ber
A_1-\tilde A_1
&=&-{1\over 16\pi^2}\int_{-\infty}^\infty ds \int_{-\infty}^\infty dt\,
{1\over (s-t)^2}\left(-2{s^2\over s^2+1}-2{t^2\over t^2+1}
+4{st(st+1)\over (s^2+1)(t^2+1)}\right)
\nonumber\\
&=&{1\over 8\pi^2}\int_{-\infty}^\infty ds \int_{-\infty}^\infty dt\,
{1\over (s^2+1)(t^2+1)}
={1\over 8} \ .
\eerl{opp}

Finally, we note that the calculation of the quadratic piece of the
Wilson loop in the case of the circle
and the straight line, which are related by an inversion through the
origin, is easy to do directly.
For the straight line we automatically get zero, since   for a straight line
\beq
\dot  x_1\cdot \dot  x_2 -|\dot  x_1||\dot  x_2| =0 \,.
\eeq
     Thus for a straight line the sum of the gluon and scalar propagators
vanishes. The reason for this triviality
is the BPS nature of our Wilson loop, which for a straight line
ensures that there are no contributions
to any order in $\lambda$.
In the case of the circle the propagators do not cancel, but their
sum is a constant, since (for $|x|=|\dot x|=1$)
\beq
(x_1-x_2)^2 = -2(\dot  x_1\cdot \dot  x_2 -|\dot  x_1||\dot  x_2|) \,.
\eeq

Explicitly, for the circle in (\ref{crc})
\beq
\vev{W}=  \int_0^{2\pi} ds dt\, {\lambda \over16\pi^2 }
{-\dot x(t) \cdot \dot x(s) + |\dot  x(t)||\dot  x(s)|
\over (x(t)-x(s))^2  } =
\int_0^{2\pi} ds dt\,{\lambda \over 16 \pi^2 }{1\over 2} ={\lambda\over 8}
\,.
\eeql{sum}

So we have learned that to quadratic order the difference between the
Wilson loop along  an open line $\tilde C$ and
along the closed contour $C$ gotten by an inversion through the origin is
\beq
\vev{\tilde W_{\tilde C}}-\vev{W_C} = -\lambda/8 \, .
\eeql{qq}
In the case of the straight line $\vev{\tilde W_{\tilde C}}=0$ and
$\vev{W_C} =\lambda/8$.
In the following we shall generalize the evaluation of the circle
and the relationship  (\ref{qq}) to all orders in $\lambda$.

\subsection{The circle to all orders}

It is simple to generalize the calculation of the circle  to
arbitrary order  in
perturbation theory.  This is because the circle is related by an
inversion to the straight line,
and the straight line receives no corrections to any order (since it is BPS).
So we start with a straight line contour $C$ (say $x(s)={1\over 2}(1,
\tan(s/2))$). The Wilson loop
along this contour is identically equal to one because of
supersymmetry. We saw this explicitly
to leading order, but  the triviality holds to all orders.  When we
perform the inversion
we will get the Wilson loop along the circle, expressed, diagram by diagram,
in terms of the diagrams for the straight line loop with the gluon
propagators modified according to
(\ref{transf}). Of course, in addition to propagators  and vertices
involving the scalars and gluons
we will also have to include  ghosts---however these, like the
scalars, transform covariantly under
the inversion.

The modifications of the gluon propagators is of the form of a
gauge transformation.   Where it not for the fact this   gauge  transformation
is singular it would have no effect on any of the diagrams of a given
order-- the boundary terms that
one would encounter upon integrating these total derivatives by parts
would cancel order by order.   This is the regular statement of gauge
invariance of the perturbative expansion. Indeed in our case we do not even
have to worry, in making these arguments, about the usual short
distance singularities that in most
theories require regularization and renormalization since the ${\cal
N}=4$ SUSYM theory is finite
when all the diagrams of a given order are included.

However, because of the singularities that occur in the modified
propagator at the origin,
the point about which the inversion is  done, there is another
boundary contribution,
namely when  and only when both ends of a single propagator hits the
origin (or the point at infinity).
As we saw above, by introducing a cutoff $\epsilon$,
when one end of the propagator hits the origin (or the point at infinity),
the resulting modification to the propagator   is of
order $\epsilon$ unless the other end of the propagator also hit  the
origin. Thus it behaves like a one-dimensional
delta-function that contributes a finite amount when
the other end of the propagator is integrated over the loop.
Therefore when  both ends of a single propagator approach the origin (or
the point at infinity) we get a constant factor of $-1/8$ ($1/8$).

One should worry about contributions when the other end of the propagator is
on an internal vertex that approaches the origin. We think that at least
for the $\cN=4$ theory those graphs will not contribute, but we were unable
to prove that. By using the same regularization as above,
it is easy to see that the contribution, if any, would come only when all
the connected part of the diagram collapses to that point. This means that
it can be described by an interaction term in the matrix model.
An explicit calculation \cite{Erickson:2000af} shows that there is no term
of order $g^4$. It would require
a better regulator and a more careful calculation to show that the interaction
terms vanish to all orders.
The remarkable agreement between our results and the $AdS$
calculation suggest that there are no interactions. In the remainder 
of the paper we will
assume that indeed all the interaction  terms vanish, and will 
provide evidence for that
from the comparison to string  theory in $AdS$.

So, ignoring interactions,
if we integrate by parts all of the modified gluon propagators we will get
non vanishing  contributions from  single propagators that are not
attached to other parts of the diagram
when  both ends of a single propagator approach the origin (or the
point at infinity). These will yield constant
factors times the rest of the diagram, as is illustrated (for a circle) in
\figur{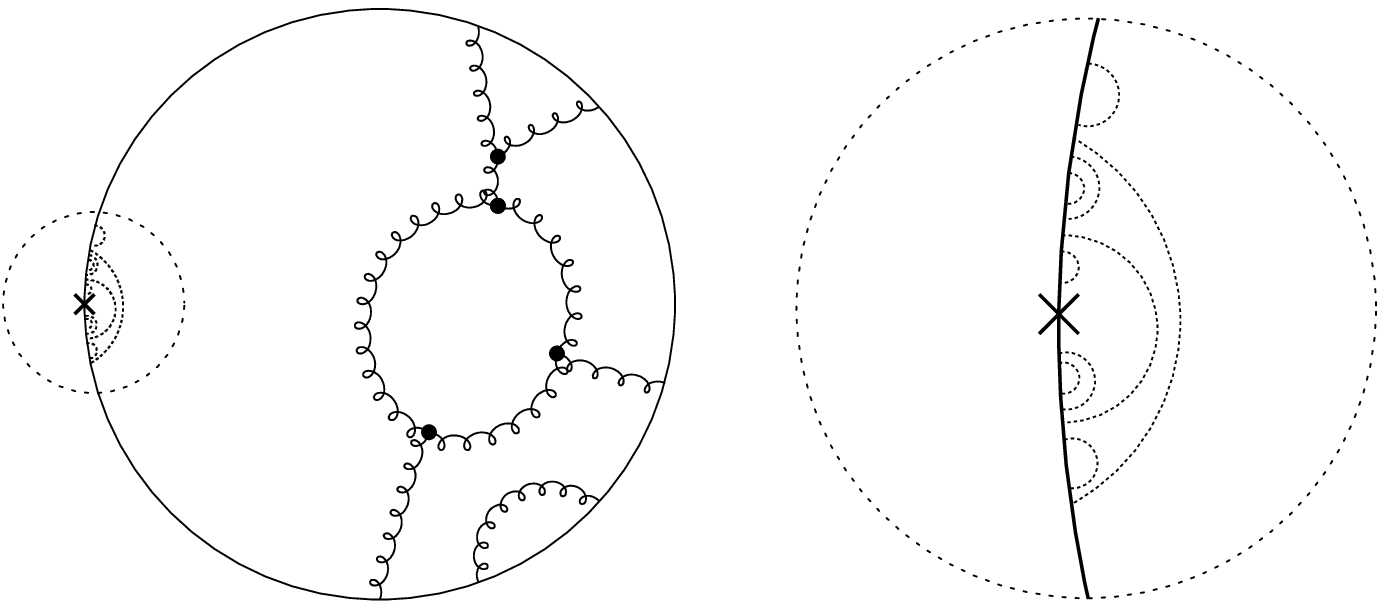}{a. To go from a straight line to a circle one should include
diagrams with some gluon propagators replaced by the total derivatives
(dotted lines). Those give a boundary contribution only when all of them
hit the point of inversion (marked by an x). b. Regardless
of the rest of the diagram, the anomaly is dependent only on the vicinity
of the inversion point and since it lives at one point, is given by the
matrix model expression.}
But the sum  of the rest of
the diagrams (to any given order)
vanishes in the case of the straight line. Therefore the calculation
of the straight line Wilson loop, with modified gluon propagators,
reduces to summing all graphs with just noninteracting modified gluon
propagators.
Each such modified propagator will give a factor  as in (\ref{opp}).
We simply have to add all these
terms.

Alternatively we can argue that since the sum of the ordinary gluon and scalar
propagators vanishes, we can add these as well. This then is
inverted to the Wilson loop for a circle, where
we should sum the Feynman diagrams of  a non-interacting theory of scalars
and vectors. This is
a simple calculation
to perform,  since as we have seen---in the case of the circle---the sum of the
gluon (in Feynman gauge) and scalar propagator contributions
is a constant (see \ref{sum}). Since each propagator just yields a
constant we can perform the sum, and account for the factors of $N$,
by doing the calculation in a 0-dimension field theory, namely a matrix model.
      This leads to the
expression
\beq
\vev{W_{\rm circle}}=
\vev{{1\over N}\Tr\exp(M)}
={1\over Z}\int \cD M {1\over N}\Tr\exp(M)
\exp\left(-{2N\over \lambda}\Tr M^2\right)\,.
\eeq

In the Appendix we shall show that this integral can be calculated exactly,
in  an expansion in powers of $1/N^2$.
     The result is (where $L_n^m$ is the Laguerre polynomial
$L_n^m(x)=1/n!\exp[x]x^{-m}(d/dx)^n
(\exp[-x]x^{n+m})$):
\ber
\vev{W_{\rm circle}}& \equiv & F(\lambda, N) ={1\over
N}L_{N-1}^1\left(-{\lambda / 4N}\right)
\exp\left[{\lambda / 8N}\right] \nonumber \\
&=&{2\over\sqrt\lambda}I_1\left(\sqrt\lambda\right)
+{ \lambda \over 48 N^2}I_2\left(\sqrt\lambda\right) +
{ \lambda^2 \over 1280 N^4}I_4\left(\sqrt\lambda\right)+
{ \lambda^{5\over 2} \over 9216
N^4}I_5\left(\sqrt\lambda\right)+\dots\nonumber \\
\eer

To leading order in $1/N$ we recover the result
\beq
\vev{W_{\rm
circle}}_{N=\infty}={2\over\sqrt\lambda}I_1\left(\sqrt\lambda\right)=
\sum_{n=0}^\infty {(\lambda/4)^n\over n!(n+1)!}
\,,
\eeq
in agreement with \cite{Erickson:2000af}, where the leading,
noninteracting, rainbow graphs  (the leading large $N$ graphs)
were summed.

Our result is based on a perturbative expansion, but we do not expect 
corrections due to instantons.
We found that the only 
contributions are from diagrams collapsed to the point of inversion, and 
since instantons are smooth objects, 
the singular graphs have measure zero, and will not contribute.

\subsection{Arbitrary loops}

As was explained in the preceding section, the contribution to the
circular Wilson loop can be localized near a single point. Going from the
straight line to the circle, the contribution is from the point at infinity.
Since the calculation can be pushed to one point, one would expect that
it does not depend on the shape of the curve. Indeed we will see that for any
smooth closed curve $C$ and the open curve $\tilde C$ related to it by a
conformal transformation the appropriate Wilson loops satisfy
\beq
\vev{W_C}
=\vev{{1\over N}\Tr\exp(M)}\vev{\tilde W_{\tilde C}}
=F(\lambda,N)\vev{\tilde W_{\tilde C}}\,.
\eeql{ttr}
We will prove this equation below by comparing Feynman diagrams of the two
Wilson loops.

First, we will explain one feature of (\ref{ttr}), the fact that the left hand
side has a single trace, while the right hand side has two
traces---over $\exp(M)$
and over the open Wilson loop. The reason for this factorization is that
the SUSYM fields and the matrix $M$ are independent variables. In general, for
two independent Hermitean matrices $A$ and $B$ with independent
$U(N)$ invariant
measures $\mu({A}),\tilde\mu({B})$,
\ber
\vev{{1\over N}\Tr(f(A)g(B))}
&=& \int \cD A\cD B\,\mu({A})\tilde\mu({B}){1\over N}\Tr(f(A)g(B))
\nonumber  \\
&=&\int \cD A\cD B\,\mu({A})\tilde\mu({B}){1\over N}\Tr(U^\dagger f(A)
U\, V^\dagger g(B) V)\,,
\eer
with arbitrary unitary $U,V$. Since they are independent, $W=U\, V^\dagger$
can take any value in $U(N)$, and   we can integrate over it
\beq
\int \cD A\cD B {\cD W\over {\rm Vol}[{U(N)}]}\mu({A})\tilde\mu({B})
{1\over N}\Tr(A W B W^\dagger)
=\int \cD A\cD B\mu({A})\tilde\mu({B}){1\over N^2}\Tr A\, \Tr B \,.
\eeql{arg}

Using this result
\beq
\vev{{1\over N}\Tr\left[\exp(M)
\cP\exp i\int_{\tilde C}(A_\mu \dot x^\mu+i\Phi_i|\dot x| \theta^i)dt\right]}
=\vev{{1\over N}\Tr\exp(M)}\vev{\tilde W_{\tilde C}}\,.
\eeq

The proof for a general loop is again diagrammatic, order by order in
perturbation theory. We write the loops again as
\beq
\vev{W_C}=\sum_{n=0}^\infty A_n \lambda^n\,,
\qquad
\vev{\tilde W_{\tilde C}}=\sum_{n=0}^\infty \tilde A_n \lambda^n\,.
\eeq
Let us look at a certain diagram $\Gamma$ of $W_C$ at order $g^{2n}$ which
contributes to $A_n$, and assume it has $k$ vertices on the Wilson loop.

There is a similar diagram $\tilde\Gamma$ contributing to the coefficient
$\tilde A_n$ of $\tilde W_{\tilde C}$. Those two diagrams are not equal to
each other, rather $\Gamma$ is equal to $\tilde\Gamma$ if we
replace the gluon propagator by the modified propagator (\ref{transf}).
Thus $\Gamma$ is equal to $\tilde\Gamma$ plus total derivatives. See
\figur{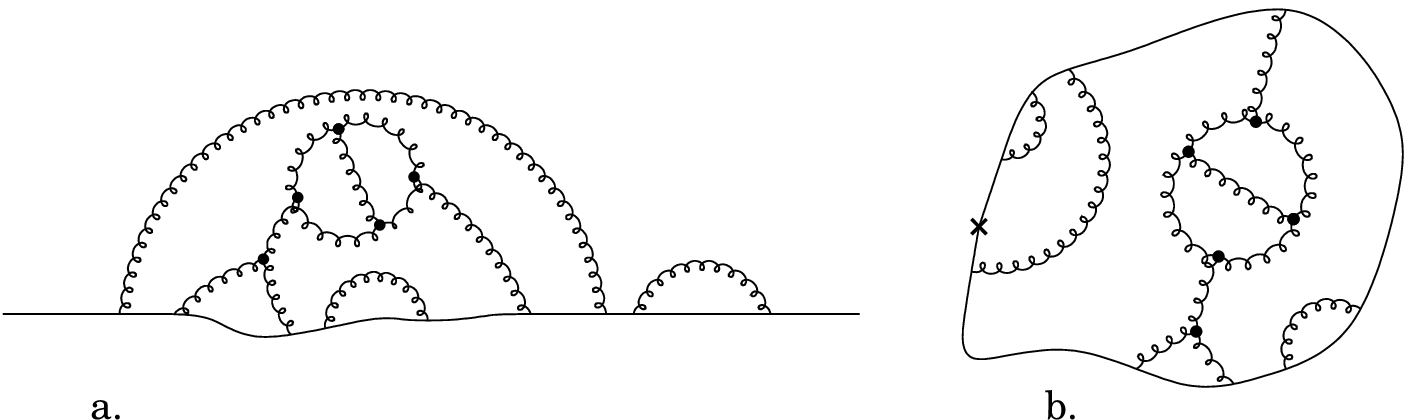}{Two graphs contributing
(a) to the open Wilson loop $\vev{\tilde W_{\tilde C}}$ and (b)
to the closed loop $\vev{W_C}$. The curves are related by a conformal
transformation, and the two diagrams differ by total derivatives}

Exactly as in the case of the circle, the total derivatives terms will cancel
unless they hit the origin. When one end hits this point the resulting
expression is proportional to a one dimensional delta-function, forcing the
other end to the origin.

\figin{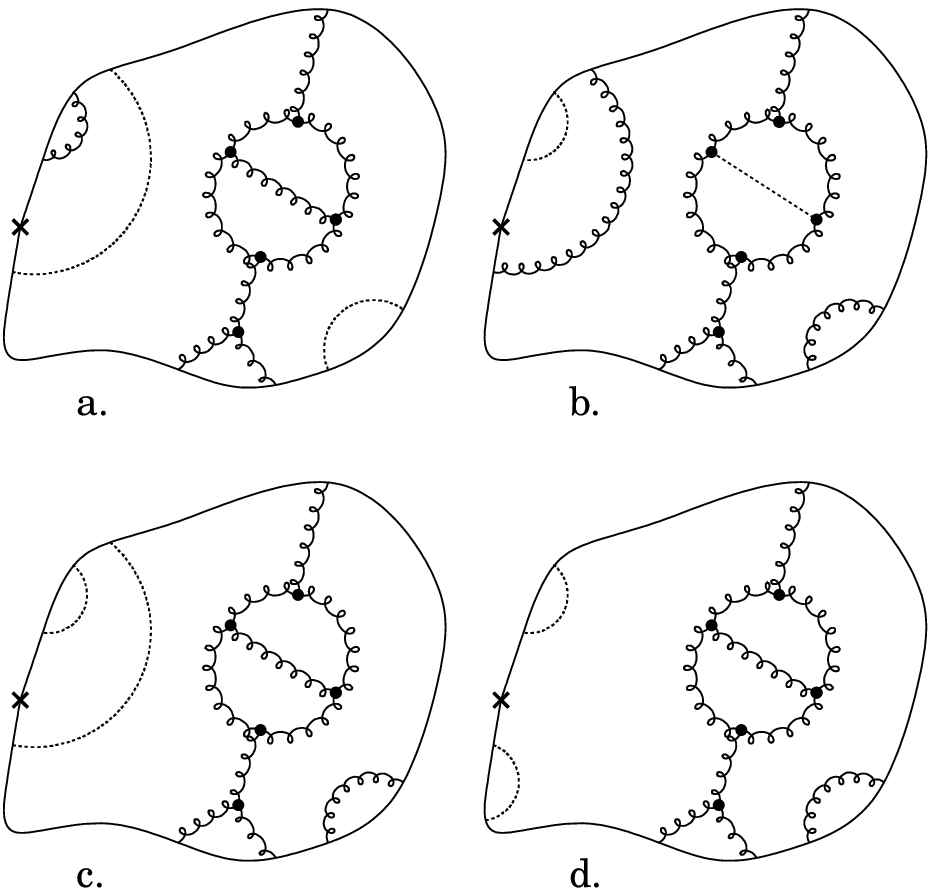}{We show here some diagrams one gets by replacing gluon
propagators by total derivatives (dotted lines). Diagrams (a) and (b) will
not contribute, since not all the total derivatives hit the inversion point.
(c) does contribute, since all the total derivatives can hit the origin.
One gets diagram (d) by doing the same procedure to a sightly different
graph. Summing (c), (d) and a few other such graphs gives the matrix model
expression at order $\lambda^2$ times the rest of the diagram.}

So considering $\tilde\Gamma$ with $l$ boundary to boundary propagators
replaced by the total derivatives will give a contribution from the singular
point times the rest of the diagram $\tilde\Gamma'$, as in Fig. \ref{fig3}c.
We find the same sub diagram $\tilde\Gamma'$ by replacing propagators by
total derivatives in other graphs, as illustrated in Fig. \ref{fig3}d.

Summing all of them we see that the total derivative contribution is exactly
the matrix model expression as before. From the example of the circle we
know that $l$ total derivatives give the same as the insertion of
${1\over(2l)!\lambda^l} M^{2l}$. Since there is only one trace, this
should be taken as a
matrix multiplying the rest of the diagram. But by the argument above
(\ref{arg}), the trace breaks in two. Therefore we see that $A_n$ is
equal to $\tilde A_n$ plus matrix model corrections
\beq
A_n
=\sum_{l=0}^n \vev{{1\over N(2l)!\lambda^l}\Tr M^{2l}} \tilde A_{n-l}
\eeq
Therefore
\beq
\vev{W_C}
=\sum_{n=0}^\infty A_n\lambda^n
=\sum_{n=0}^\infty \sum_{l=0}^n \vev{{1\over N(2l)!}\Tr M^{2l}}
\tilde A_{n-l}\lambda^{n-l}
=\vev{{1\over N}\Tr \exp(M)}\vev{\tilde W_{\tilde C}}\,.
\eeq

The crucial point in the proof is that the total derivative part of the
graphs (the matrix model) totally decouple from the rest of the graph. The
total derivatives live within an infinitesimal distance from the origin. It
is a set of measure zero for any other part of the graph to be in that
vicinity, and since the loop is smooth, and the theory is finite, this set
of measure zero does not contribute.

The above argument is true for all $N$, not just planar graphs. Again, one
has to note that if the matrix model part has genus $p$ and the rest of the
graph is at genus $q$, the total genus is $p+q$, since those two graphs are
totally separated. Also, we assumed here that the matrix model is quadratic,
but the statement would be correct regardless of that. Even if the
interactions don't vanish, to get a contribution, the entire part of the
diagram with interaction has to collapse to the singular point. It would
still give a matrix model contribution times the rest of the graph.

%%%%%%%%%%%%%%%%%%%%%%%%%%%%%%%%%%%%%%%%%%%
\mysection{The Comparison with String Theory}
%%%%%%%%%%%%%%%%%%%%%%%%%%%%%%%%%%%%%%%%%%%

The $AdS$/CFT \cite{Maldacena:1998re} correspondence allows one to calculate
the expectation value of Wilson loops in $\cN=4$  SUSYM
for  large $\lambda$ from minimal surfaces in $AdS$ space
\cite{Rey:1998ik,Maldacena:1998im}. We will now compare our
calculation of the ratio of
Wilson loops that are related by inversion, as well as  the exact
expression for a circular Wilson loop,
to string theory calculations.

We have shown that a Wilson loop, $W_C$,  along a closed contour $C$
passing through the origin, is related to a Wilson loop, $\tilde
W_{\tilde C} $,  along  the open line, $\tilde C$, gotten by
inverting the contour through the origin,
by:
\beq
\vev{{W}_{  C}}
= F(\lambda, N)\vev{{\tilde W}_{\tilde C}}\,.
\eeql{rel}
We would like to prove the same statement from  string theory.
A complete proof is beyond our capabilities, since the calculational
tools for  string perturbation theory in $AdS_5$
     are still undeveloped. However, we are able to    give strong
evidence from string theory for this relationship.
to leading order in $1/\lambda=(l_s/R)^4$, and to all orders in the
string coupling, $g_s=\lambda/(4\pi N)$ , for arbitrary smooth loops!

\subsection{Circular loops}

For circular loops we can perform a   precise test of
the $AdS$/CFT correspondence,
since we have derived an exact expression for the circular Wilson loop
for all $\lambda$ and $N$. In string theory, to a given order in
$1/N^2$, we expect  that the Wilson loop should be given by
\beq
\vev{W_{\rm circle}}_p = { 1\over N^{2p }}e^{-S_p}  f_p(\lambda) \,,
\eeq
where $S_p$ is the  action for a minimal surface ending on the circle
with $p$ handles
and $f(\lambda)$ would be calculated by evaluating the fluctuations
about the minimal surface in powers
of $\alpha'$ (or $l_s/R$, or equivalently $1/\lambda^{1/4}$).

The minimal area surface  to leading order in $1/N^2$ can be
constructed analytically
and yields $S_0= -\sqrt{\lambda}$, it is a smooth, geodesic surface.
To higher order in
$1/N^2$ we need to find the minimal area surface with handles. It is
intuitively obvious that
the best we can do is to attach degenerate handles that have no area to the
above surface. This is not a
smooth surface, but it is the limit of smooth surfaces and has the
minimal possible area.\footnote
{We have been assured by M. Freedman that when the boundary is
a round circle this can be proven by a standard projection argument.}
If this  is the case then $S_p=S_0=-\sqrt{\lambda}$.

To do better than this one would need to evaluate the stringy
fluctuations about the minimal
surface,  in an expansion in $\alpha'$. This is beyond our
capabilities. However, we can
determine the overall power of the inverse coupling, $1/(l_s/R)$ that
multiplies $e^{-S}$. We claim that
\beq
\vev{W_{\rm circle}}_p^{\rm string}
\propto
{1\over N^{2p}}
{\lambda^{{6p-3\over 4}}\over p!}
e^{\sqrt{\lambda}}
\left[\,1 + O\left({1\over \sqrt{\lambda}}\right)\,\right] \,.
\eeql{cst}
The factor of $1/p!$ arises since the handles are indistinguishable.
We give two arguments for the power of $\lambda$ in this expression. The
string coupling is $g_s^2\sim \lambda^2/N^2$, but in addition one has to be
careful the contribution of zero modes. The dimension of the moduli space
of surfaces of genus $p$ with one boundary is $6p-3$. Since the relevant
surfaces are degenerate we have to impose two real constraints for each
handle, in addition to the overall 3. Each constraint gives a power of
$\lambda^{-1/4}$, from the correct normalization of the zero modes. This
gives
\beq
\left({\lambda\over N}\right)^{2p}
\to{\lambda^{6p-3\over4}\over N^{2p}}\,.
\eeq

An equivalent calculation comes from the low energy effective supergravity,
the degenerate handles are the same as the exchange of supergravity modes.
In \cite{Berenstein:1999ij} the exchange of fields between two widely
separated surfaces was calculated. One can redo their calculation for
the case at hand, the self interaction of the surface ending on a circle.
In their case the coupling of the Kaluza-Klein modes is proportional to
$1/N^2$ and the integration over each of the surfaces gives a measure
factor of $\sqrt\lambda$.

Therefore the result for well separated surfaces was proportional to
$\lambda/N^2$. For calculating the self interaction of a single surface we
have to use the propagator at short distances, which, in 5 dimensions,
has a cubic divergence. Integrating over the surface leaves a linear
divergence, which should be cut off at the string scale, giving an
extra factor of $R/l_s\sim\lambda^{1/4}$. In addition we should sum over all
the KK modes, again imposing a cutoff---the angular momentum cannot exceed
$R/l_s$. This gives the final result $\lambda^{3/2}/ N^2$ for each
handle.

This power of $\lambda$ is also confirmed by the S-duality argument in the
following section.

We can now compare this with the gauge theory result, $\vev{W_{\rm
circle}}=F(\lambda,N)$.
In Appendix B we examine the large $\lambda$ behavior of the $1/N^2$
expansion of
$F(\lambda,N)$. We show that, order by order in the $1/N^2$
expansion, this function behaves, for large $\lambda$, as:
\beq
\vev{W_{\rm circle}}^{\rm gauge}= F[\lambda,N]=
\sum_p    {1\over N^{2p}}
{e^{\sqrt{\lambda}}\over  p!} \sqrt{2\over{\pi}}
     {{\lambda^{6p-3\over 4}} \over 96^p}   \left[ 1 -
{3(12p^2+8p+5)\over 40 \sqrt{\lambda}}+  O\left({1\over
    {\lambda}} \right)\right] \, .
\eeql{mmr}
Thus we find precise agreement with the string theory calculation,
order by order in $1/N^2$, to leading order in $1/\lambda\, $!

\subsection{S-duality}

Another very strong test of this expression comes from checking its region
of validity.\footnote{We thank Sunny Itzhaki for suggesting this calculation.}
Clearly both the $AdS$ expression (\ref{cst}) and the matrix model result
(\ref{mmr}) are valid for $\lambda\gg1$. If we ignore the $1/\sqrt\lambda$
correction the matrix model gives
\beq
\vev{W_{\rm circle}}^{\rm gauge}\sim
\sqrt{2\over{\pi}}\lambda^{-{3\over 4}}
\exp\left(\sqrt{\lambda}
+{\lambda^{3\over 2} \over 96 N^2}\right)\,.
\eeq
Thus the approximation $\vev{W_{\rm circle}}\sim \exp\sqrt\lambda$ is valid
as long as $1\ll\lambda\ll N^2$. The $AdS$ expression is valid only for
$\lambda\ll N$, or else string theory is strongly coupled. For $\lambda\gg N$
we should perform an S-duality transformation. Under
S-duality  the Wilson loop turns into an 't Hooft  loop of
the dual theory described by a
D1-brane. The action for this  configuration is given in terms of
the dual couplings $\tilde
g_s=1/g_s$  and $\tilde \lambda = \lambda/g_s^2$
\beq
\vev{W_{\rm circle}}^{\rm dual\ string}
\sim \exp {\sqrt{\tilde\lambda}\over \tilde g_s}
=\exp\sqrt\lambda\,.
\eeq
So the dual D1-brane has the same action as the original fundamental string.
This dual description is valid as long as $\tilde\lambda\gg1$, or
$\lambda\ll N^2$. We see, therefore, that the range of validity of the
two calculations is identical!

This can be regarded as another test of the matrix model expression, and in
particular the power of $\lambda$ accompanying the $1/N^2$ corrections. But
it should also be considered a test of S-duality in $\cN=4$ SUSYM. The
matrix model is valid for all values of $g$, and with the replacement
$g\to 4\pi/g$ it gives the value of the 't Hooft loop, which is confirmed
by the $AdS$ calculation.

\subsection{Arbitrary loops}

This story can be generalized, to some extent, to arbitrary loops.
Indeed, a version of this statement for large $\lambda$
and to lowest order in $g_s$ was made in a footnote in \cite{Drukker:2000ep}.
As shown in  \cite{Drukker:1999zq}, the  expectation
value of the Wilson loop to leading order in the $\alpha' $ expansion,
is
\beq
    \vev{W} \propto e^{-S} \,,
\eeq
where the action, $S$, is a Legendre transform of the area
of the surface in $AdS_5$  whose boundary is the loop contour.
The  Legendre  transform  removes (for a smooth loop)
the divergence in the area. For smooth loops the Legendre transform is equal
(asymptotically) to the extrinsic curvature of the boundary $\kappa $. Then we
can use the Gauss-Bonnet theorem to write the action for the minimal area as:
\beq
S
={\sqrt{\lambda}\over2\pi}
\left[\int d^2\sigma\sqrt{g}-\int d\tau\sqrt\gamma\kappa\right]
={\sqrt{\lambda}\over2\pi}\int d^2\sigma\sqrt{g}
\left(1+{1\over 2}R^{(2)}\right)
-\sqrt\lambda\chi\,,
\eeq
where $R^{(2)}$ is the induced metric and $\chi$ the Euler number of the
surface (given by this integral expression). It is easy to see that
$R^{(2)}$ approaches $-2$ near the boundary of $AdS$, so the integral on the
right hand side is manifestly convergent.

The action  integral is invariant under isometries of $AdS$ including
conformal transformations. Since it is manifestly convergent, it is invariant
also if the  conformal transformation takes a point from finite
distance to infinity, or vice versa. What does change in the latter
case is the topology of the surface. The Euler number is one for the disc,
the appropriate world sheet for a closed Wilson loop $W_C$. But for the open
Wilson loop $\tilde W_{\tilde C}$ the world sheet is the half plane with
Euler number zero. Therefore
\beq
\vev{W_C}
=\exp\left(\sqrt\lambda\right)\vev{\tilde W_{\tilde C}}\,,
\eeq

In fact this statement can be generalized to any order in the string
coupling, or the $1/N^2$, expansion. This is clearly the case if the
minimal surface at higher genus is gotten by adding degenerate handles
to the surface of lower genus---the handles do not change the action.
But the proof does not require this assumption.
To order $1/N^{2p}$ the relevant
surface bounding the closed contour is
topologically a disk with $p$ handles, for which  $\chi=1-2p$, whereas
the  surface bounding the open contour is a half plane with $p$
handles, for which  $\chi=-2p$.
Consequently, to any order in $1/N^2$ and
for large $\lambda$, we expect from string theory that:
\beq
\vev{W_C}
=\exp\left(\sqrt\lambda ( 2p+1-2p)\right)\vev{\tilde W_{\tilde C}}
=\exp\left(\sqrt\lambda\right)\vev{\tilde W_{\tilde C}}\,,
\eeq
This is precisely what we find in the gauge theory from (\ref{rel}),
using the result proved in Appendix B that,
to any  order in $1/N^2$
\beq
F(\lambda, N) \sim e^{\sqrt{\lambda}}\,,
\eeq
for large $\lambda$. Thus (\ref{rel}) is true to leading order in $1/\lambda$.

Understanding the $1/\sqrt\lambda$ corrections is more difficult, since
we cannot even calculate the expectation value of an arbitrary open
loop. Still,
the string theory argument leading to (\ref{cst}) is general and should
apply to any closed curve (as long as there are no new smooth classical
solutions at high genus). Therefore we might expect that:
\beq
\vev{W_C}_p^{\rm string}
\propto
{1\over N^{2p}}
{\lambda^{{3p\over 2}-{3\over 4}}\over p!}
e^{-S}
\left[\,1 + O\left({1\over \sqrt{\lambda}}\right)\,\right] \,.
\eeq

This might look surprising, given that the corresponding open loop
$\vev{\tilde W_{\tilde C}}$ is not one. The reason that it works is that the
   open loop asymptotes to a straight line, so it differs significantly
from the BPS straight line
only over a compact region. We can expect that the
leading behavior of the asymptotically straight line and the true straight line
would be the same. If a genus $p$ surface is gotten by adding $p$
degenerate handles, then there is a large probability that they will
be attached
within  the asymptotically straight part of the world sheet, where
they will not
contribute because of supersymmetry. Therefore, for most of the volume of the
moduli space, we will get no enhancement and we might conjecture that
to order $1/N^{2p}$:
\beq
\vev{\tilde W_{\tilde C}}_p^{\rm string}
\propto
{1\over N^{2p}}
e^{-S-\sqrt{\lambda}}
\left[\,1 + O\left({1\over \sqrt{\lambda}}\right)\,\right] \,.
\eeq
Under these assumptions, the relation derived from the gauge
theory,(\ref{rel}),
will agree with the string theory to all orders in
$1/N^2$ for large $\lambda$, since
\beq\vev{W_C}^{\rm string}
\propto
\sum_p{1\over N^{2p}}
{\lambda^{{6p-3\over 4 }}\over p!}e^{-S}
\sim
\left[\sum_p{e^{\lambda}\over N^{2p}}
{\lambda^{{6p-3\over 4 }}\over p!} \right]
\left[\sum_q {1\over N^{2q}}
e^{-S-\sqrt{\lambda}}\right] \,.
\eeq
\mysection{Multiply wound loops}

The above considerations can be extended to multiply wound Wilson 
lines or loops.
Consider, for example, a Wilson loop consisting of two coincident 
circles. These can be tied
together so that the loop winds twice around a circle, or traced 
independently.
Under an inversion  through a point on the circle they go into two 
coincident parallel
straight lines, which are  BPS and thus trivial. By the same 
arguments that we have
presented above  the evaluation of the multiply wound loops can be 
expressed in terms of
the matrix model.

Consider first two circles on top of each other. If the untraced 
Wilson loop around the circle
is denoted by ${\cal W}$, so that the ordinary Wilson loop traced 
around one circle is
$  W_1= 1/N\vev{\Tr\, {\cal  W}}$, then the two options for 
connecting the circles correspond to
$ W_{2}=1/N \,\vev { \Tr\, {\cal  W}^2  } $ and to $  W_{1,1}=
1/N^2 \,\vev {\left(\Tr\, {\cal  W}\right)^2 }$ respectively.  In 
terms of the matrix model it is clear
that
\ber
  W_{2}&=&{1\over N}\vev { {\Tr} \exp(2M)  }
  \, ,\nonumber \\
  W_{1,1}&=&{1\over N^2}\vev {\left[{\Tr} \exp(M)\right]^2} \, .
\eer

    The first case, that of doubly wound loop, is very simple. Scaling 
$M\to M/2$, we see that
  the result is the same as the single     circle with $\lambda\to 
4\lambda $, thus
  \beq
   W_{2}(\lambda,N)=W_1(4\lambda,N)= {1\over 
N}L_{N-1}^1\left(-{\lambda /N}\right)
\exp\left[{\lambda /2N}\right].
  \eeq
  In the case of the squared singly wound loop we follow the same 
steps as in Appendix A:
  \ber
   W_{1,1}&=&{1\over Z} \int {\cal D} M \left[{1\over N }{\Tr} 
e^M\right]^2 e^{-{2N\over \lambda}
  \Tr M^2}
\nonumber
\\
  &=& {1\over Z} \int dm_i \Delta^2(m_i) \left[{1\over N }\sum_i 
e^{m_i}\right]^2 e^{-{2N\over
\lambda}
  \sum m_i^2 } \nonumber
\\
  &=&  {1\over Z'} \int dm_i \Delta^2(m_i)   e^{- \sum m_i^2 }
  \left[ {1\over N }e^{2m_1\sqrt{\lambda\over 2N}} +{N-1\over 
N}e^{(m_1+m_2)\sqrt{\lambda\over
2N}}\right]   \eer
The first integral  is, up to a factor of $1/N$, the same as $W_2$. The second
can be evaluated by expressing, as in Appendix A, the Vandermonde 
determinant, $\Delta^2(m_i)$,
in terms of Hermite polynomials, as
\beq
{1\over N^2} \int dm dm'\sum_{i,j=0}^{N-1}\left[P_i^2(m)P_j^2(m')-P_i 
(m)P_j (m)
P_i (m')P_j(m')\right]e^{-(m^2+m'^2)+\sqrt{\lambda\over 2N}(m+m')}
\, .
\eeq
  The  above   integrals can then be done, with the final result being
  \beq
  W_{1,1} ={1\over N} W_2 
+\left(1-{1\over N} \right)(W_1)^2-{2\over N^2}e^{\lambda\over
4N}\sum_{i=1}^{N-1}\sum_{j=0}^{i-1}\left[L^{i-j}_j\left(-{\lambda\over
4N}\right)\right]^2 \, .
  \eeql{mul}

  One of the sums in  (\ref{mul})  can easily be done and the result 
compared with string theory
  for large $\lambda$. It is trivial to reproduce the correct 
semiclassical action and it would be
interesting to try to account for the factors of $\lambda$ as well. 
A similar analysis can be carried
out,
with increased complication, for loops wound any number of times 
around the circle. 
In fact, it does not have to be the exact same circle, 
one gets the same result from arbitrary loops that 
are tangent to each other at one point. Under an inversion around 
the common point they are mapped to a collection of parallel lines 
which is also a BPS configuration.

These Wilson loops correspond to the
most general observables  of the matrix model,
\beq  
%W_{i_1,i_2,\dots i_n} \equiv \vev {  \Tr (M^{i_1})\Tr 
%(M^{i_2})\dots \Tr (M^{i_n}) } \,  ,
W_{i_1,i_2,\dots i_n} \equiv \vev {  \Tr \exp(i_1 M)\ 
\Tr \exp (i_2 M)\cdots \Tr \exp(i_nM) } \,  ,
\eeq
and can be used, following the discussion in (\cite{grotay}, \cite{groog}),
to evaluate the expectation values of Wilson loops in definite 
representations of $U(N)$.
We  postpone this analysis for elsewhere.

\mysection{Conclusions}

In this paper we have extended, generalized and outlined a proof for the
result of Erickson, Semenoff and  Zarembo
\cite{Erickson:2000af} on the value of the circular Wilson loop
in ${\cal N}=4$ SUSYM. We showed  that the expectation value of a
circular  BPS-Wilson loop in ${\cal N}=4$ SUSYM is determined by an anomaly
in the conformal transformation that relates the circular and
straight-line loops.
As such it can be calculated exactly, to all orders in a $1/N^2$
expansion and to
all orders in $g^2N$. A similar relation was derived
    between
the expectation value of   any closed smooth Wilson loop  and  the loop
related to it by an inversion
that takes a point along the loop to infinity. Using the AdS/CFT
duality, this result  yielded a
prediction of  the value of  the string amplitude with a circular
boundary to all orders in $\alpha'$ and to all orders
in $g_s$.  We then compared this result with string theory,
and found that the gauge theory  calculation, for  large $g^2N$ and to
{\it all} orders in the $1/N^2$
expansion does agree
with the leading   string theory  calculation, to all orders in $g_s$
and to lowest order in $\alpha'$.

We proved that the anomaly is given by a matrix model, but we leave for
future work to complete the proof that all interactions vanish and  the
matrix model is indeed quadratic. The agreement with the AdS calculation is a
very strong indication that the quadratic matrix model is correct, at least
for the $\cN=4$ theory. In principle the anomaly in other conformal field
theories could be described by a more complicated matrix model.

This  agreement is remarkable. It is a test of the AdS/CFT correspondence
in the regime of strong gauge coupling (small $\alpha'$) and to all
orders in $1/N^2$, the string coupling. The result even extends to the S-dual
region where the fundamental string is replaced by a D1-brane. This gives
strong evidence for the validity of the conjectured AdS/CFT correspondence.

All the calculations in this paper were done for gauge group $U(N)$, but
the generalization to $SU(N)$ is trivial. We write the Hermitean matrix $M$
as the sum of a traceless part and the trace times the unit matrix
$M=M'+m I_N$. Then
\beq
\vev{{1\over N}\Tr\exp M}_{U(N)}
=\exp\left({\lambda\over 8N^2}\right)
\vev{{1\over N}\Tr\exp M'}_{SU(N)}\,.
\eeq
In string theory the difference between $SU(N)$ and $U(N)$ corresponds
to the inclusion of some fields
that do not have local dynamics, but can be gauged to   infinity. In any
case the difference is subleading in both $N$ and $\lambda$, so it has no
consequence on our discussion of the leading behavior for large $\lambda$,
order by order in $1/N^2$.

It would be very interesting to try to understand the  $\alpha'$ corrections
to the minimal surface calculation in AdS, in order to compare our
exact result with string theory.
Consider the leading $N=\infty$ prediction for the circular Wilson loop. Using
the asymptotic
expansion of the Bessel function, we can write the expectation value of the
circular loop as
:\beq
\vev{W_{\rm circle}}^{\rm gauge}
=\sqrt{2\over\pi}{e^{\sqrt\lambda}\over\lambda^{3/4}}
\sum_{k=0}^\infty \left({-1\over2\sqrt\lambda}\right)^k
{\Gamma({3\over2}+k)\over\Gamma({3\over2}-k)}
-i\sqrt{2\over\pi}{e^{-\sqrt\lambda}\over\lambda^{3/4}}
\sum_{k=0}^\infty \left({1\over2\sqrt\lambda}\right)^k
{\Gamma({3\over2}+k)\over\Gamma({3\over2}-k)}\,.\\
\eeql{asy}
The challenge is to reproduce, in an $\alpha'$ expansion, the
asymptotic expansion
given
in (\ref{asy}). Note that this asymptotic expansion is not Borel summable.
The terms behave as $\left({k\over2\sqrt\lambda}\right)^k  $, to order $k$.
it would be interesting to understand this from the point of view of the
world sheet theory. The non-Borel summability,
as well as the second term in (\ref{asy}), might indicate that there
is an instanton contribution
to the world sheet amplitude.%, perhaps arising from some other space-time.

Finally, it  is interesting that the string theory with a circular
boundary is described
by the Hermitean matrix model. This model  is related to non-critical
string theory
with $c=-2$ \cite{Kostov}. Here it yields a particularly simple
observable of the
critical superstring theory
in the AdS background. It is conceivable that one could derive the matrix model
representation of the string amplitude directly, without having to
use the duality to gauge theory.

\appendix
\mysection{Matrix Model Calculation}

We wish to evaluate
\beq
\vev{{1\over N}\Tr\exp(M)}
={1\over Z}\int \cD M {1\over N}\Tr\exp(M)
\exp\left(-{2N\over \lambda}\Tr M^2\right)\,.
\eeq
First, we do the angular integrations, to rewrite the integral
in terms of the eigenvalues of $M$:
\ber
\vev{{1\over N}\Tr\exp(M)}
&=&{1\over Z}\int \prod d m_i \Delta^2(m_i) {1\over N}\sum e^{m_i}
\exp[ -{2N\over \lambda}\sum m_i^2]\nonumber\\
     &=&{1\over Z}\int \prod d m_i \Delta^2(m_i)
\exp\left[\sqrt{\lambda\over 2N}{m_1}\right]
\exp[ -\sum m_i^2]
     \,.
\eer
\noindent where $\Delta(m_i) = \prod_{i<j}(m_i-m_j) =\det[\{m_i^{j-1}\}]$
is the Vandermonde determinant, and we have rescaled the $m_i$ absorbing
the normalization into $Z$.

Now we use the standard trick, \cite{Mehta}, of rewriting
this determinant in terms of orthogonal polynomials.
It is clear that, in evaluating the determinant
of the matrix $\{m_i^{j-1}\}$, we can replace the row $m_i^{j-1}$, for a given
$i$, by any polynomial in $m_i$ of rank $j-1$, that starts with $m_i^{j-1}$.
We can choose these polynomials to be orthonormal with respect  to the measure
$\int dm \exp[-  m^2]$, thus rendering
the resulting integrals easy. The appropriate polynomials are
proportional to the
the Hermite polynomials
\beq
H_n(x)= e^{ x^2}\left(-{d\over dx}\right)^ne^{ -x^2}, \quad
\int_{-\infty}^{ \infty} dx  e^{ -x^2} H_n(x)H_m(x) =\delta_{nm} 2^n
n!\sqrt{\pi} \, .
\eeq
So we choose the polynomials to be the orthonormalized Hermite polynomials
(with respect to the measure $dx  \exp( -x^2)$)
\beq
P_n(x)\equiv {H_n(x)\over\sqrt{ 2^nn!\sqrt{\pi}}},
\eeq
and write $\Delta(m_i)\propto \det[\{P_{j-1}(m_i)\}]$,
again absorbing the normalization into $Z$. The integrals over
$m_i,\,  i=2\dots N $,
can easily be done leaving us with:
\beq
\vev{{1\over N}\Tr\exp(M)}= {1\over N}\int_{-\infty}^{ \infty}
      dm \sum_{j=0}^{N-1}P_j(m)^2 \exp\left[-m^2+\sqrt{\lambda  \over 2N}
m\right]\, .
\eeq
Using the integral,
\beq
\int_{-\infty}^{ \infty} dm P_j(m)^2
\exp\left[-\left(m -{\sqrt{ \lambda\over 8N}}\right)^2\right]=
L_j\left(-{\lambda /4 N}\right) \, ,
\eeq
where $L_n^m$ is the Laguerre polynomial $L_n^m(x)=1/n!\exp[x]x^{-m}(d/dx)^n
(\exp[-x]x^{n+m})$, ($L^0_n=L_n$),
we obtain:
\ber
\vev{{1\over N}\Tr\exp(M)}&=&{1\over N} \sum_{j=0}^{N-1}
L_j\left(-{\lambda / 4N}\right)\exp\left[{\lambda / 8N}\right]
={1\over N}L_{N-1}^1\left(-{\lambda / 4N}\right)
\exp\left[{\lambda / 8N}\right]\nonumber \\
& =&{2e^{-\lambda/8N}\over N!\sqrt{\lambda/N}}
\int_0^\infty dt\, e^{-t} t^{N-\half}I_1\left(\sqrt{t\lambda/N}\right)\, .
\eerl{result}

In order to exhibit the ${1\over N}$ expansion we
write  (\ref{result}) as a power series in $\lambda$
\ber
\vev {{1\over N}\Tr\exp(M)} &=&   \exp\left[{\lambda / 8N}\right]
\sum_{k=0}^{N-1} {N \choose k+1}
      {\lambda^k\over 4^k  N^{k+1} k!}\nonumber \\
&=&  \sum_{n=0}^{\infty} {\lambda^n\over 4^n n!(n+1)!}B(n,N)\,  ,
\nonumber \\
\eerl{expan}
where
\beq
B(n,N)\equiv
\sum_{k=0}^n {n!(n+1)!2^{k-n}(N-1)!\over
k!(k+1)!(n-k)!(N-1-k)!N^n}={(n+1)!\over (2N)^n}
F(-n,1-N;2;2)  \,,
\eeql{bn}
and $F$ is the hypergeometric function ($F(\alpha, \beta;\gamma ;z)=
1 +{\alpha  \beta\over\gamma \cdot 1}z+
{\alpha(\alpha+1)  \beta(\beta+1)\over\gamma(\gamma+1) \cdot 2!}z^2
+\dots   $).
$B(n,N)$  can easily be expanded
in a power series in $1/N^2$ to yield
\ber
B(n,N)&=& 1 +{n(n^2-1)\over 12N^2} +{(n+1)!\over (n-4)!}{(5n-2)\over 1440N^4}
+ {(n+1)!\over (n-6)!}{(35n^2-77n+12)\over 2^73^45\cdot 7N^6} +
\nonumber \\
     &&\hskip -.5pt +{(n+1)!\over (n-8)!} {(175n^3-945n^2+1094n-72)\over
2^{11}3^5 5^27N^8} +\dots
\eer
Using the definition of the Bessel function: $ I_n(2x)=\sum_{k=0}^\infty
{x^{n+2k}\over k!(n+k)!}\,,$ we can then use
this expansion to derive the asymptotic expansion
in powers of $1/N$,
\beq
\vev{{1\over N}\Tr\exp(M)}={2\over\sqrt\lambda}I_1\left(\sqrt\lambda\right)
+{ \lambda \over 48 N^2}I_2\left(\sqrt\lambda\right) +
{ \lambda^2 \over 1280 N^4}I_4\left(\sqrt\lambda\right)+
{ \lambda^{5\over 2} \over 9216 N^4}I_5\left(\sqrt\lambda\right)+...
\eeq

\mysection{Explicit $1/N$ Expansion}

We now present a systematic $1/N^2$ expansion of $F(\lambda,N)$.
To this end we use the transformation formula,
$ F(\alpha, \beta;\gamma ;z)=(1-z)^{-\alpha}F(\alpha,
\gamma-\beta;\gamma ;z/(z-1))$,
to rewrite
\beq B(n,N)=(-)^n{(n+1)!\over (2N)^n}
F(-n,N+1;2;2)\,,
\eeql{bnp}
and then we use the Gauss recursion relation,
$$(2\alpha -\gamma-\alpha z+\beta z)F(\alpha, \beta;\gamma ;z)
+(\gamma-\alpha)F(\alpha-1, \beta;\gamma ;z)
+\alpha(z-1)F(\alpha+1, \beta;\gamma ;z)=0 \,,
$$
to derive the recursion relation:
\beq
B(n+1,N)=B(n,N)+{n(n+1)\over 4N^2}B(n-1,N)\,.
\eeql{recursion}

This recursion relation allows us to derive a systematic expansion of
$B(n,N)$ in powers of
$1/N^2$, starting with $B(0,N)=1$. It is easy to verify  from
(\ref{recursion}) that
$$B(n,N) = \sum_{k=0}^{[{n\over 2}]} {b_k(n)\over N^{2k}}
\,,$$
\noindent where $b_k(n)$ is a  polynomial in $n$ of rank $3k$. It is
also easy to
see that $b_k(n)=0$ for $n=0,1,2,\dots, 2k-1$.
We  can therefore  expand these polynomials
in  terms of the $k$ polynomials,
${(n+1)!/ (n-3k+1+i)!}$, that vanish for $n\leq 2k-1$:
\beq b_k(n)= \sum_{i=0}^{k-1}{(n+1)!\over (n-3k+1+i)!}X^i_k \,.
\eeql{bks}

To determine the $X^i_k$ we use  (\ref{recursion}) to derive:
\beq
4X^i_k ={3k-i-2\over 3k-i}X^{i-1}_{k-1}+{1\over 3k-i}X^{i}_{k-1}\,,
\eeql{rec2}
which, together with $X^0_1=1/12$,and $X^k_k=0$, can be used to
evaluate the $X^i_k$'s. In particular,
\beq
X^0_k = {1\over 12^k k!}\,\,; \quad X^1_k
={3\over 20}{1 \over  12^{k-1} (k-2)! }\,.
\eeql{xs}

The advantage of this expansion is that when we plug (\ref{bks})
into  the expression, (\ref{expan}), for $F(\lambda,N)$ the sum over
$n$, order by order in
$1/N^2$, can easily be performed  to derive:
\beq
F(\lambda,N)= {2\over \sqrt\lambda } I_{1}({\sqrt\lambda
})+\sum_{k=1}^{\infty}{1\over N^{2k}} \,\sum_{i=0}^{k-1}\, X^i_k\,
\left({\lambda\over 4}\right)^{3k-i-1\over 2} I_{3k-i-1}({\sqrt\lambda }) \,.
\eeql{finalF}

This expression can then be used to determine the large
$\lambda$ behavior of $F$, order by order in $1/N^2$,
\beq
F(\lambda,N) =
\sum_p    {1\over N^{2p}}
e^{\sqrt{\lambda}} \sqrt{2\over{\pi}}
     {{\lambda^{6p-3\over 4}} \over 96^p   p!}   \left[ 1 -
{3(12p^2+8p+5)\over 40 \sqrt{\lambda}}+  O\left({1\over
    {\lambda}} \right)\right] \, .
\eeq

\section*{Acknowledgements}

We would like to thank Michael Freedman, Sunny Itzhaki, 
Joachim Rahmfeld, Arkady Tseytlin and Ed Witten  for discussions, and  
Hirosi Ooguri for also pointing out some problems with the original 
manuscript.
The work of N.D.was supported by the DOE grant DE-FG03-84ER40168.
The work of D.J.G was supported by the NSF under  the
grants  PHY 99-07949 and PHY 97-22022.


\begin{thebibliography}{20}
%%%%%%%%%%%%%%%%%%%%%%%%%%%
\addtolength{\parskip}{-1ex}


\bibitem{Erickson:2000af}
J.~K.~Erickson, G.~W.~Semenoff and K.~Zarembo,
``Wilson Loops in $N = 4$ Supersymmetric Yang-Mills Theory,''
Nucl.\ Phys.\  {\bf B582}, 155 (2000)
[hep-th/0003055].
%%CITATION = HEP-TH 0003055;%%

\bibitem{Maldacena:1998re}
J.~Maldacena,
``The Large $N$ Limit of Superconformal Field Theories and Supergravity,''
Adv.\ Theor.\ Math.\ Phys.\  {\bf 2}, 231 (1998)
[hep-th/9711200].
%%CITATION = HEP-TH 9711200;%%

\bibitem{Rey:1998ik}
S.~Rey and J.~Yee,
``Macroscopic Strings as Heavy Quarks in Large $N$ Gauge Theory and
Anti-de Sitter Supergravity,''
hep-th/9803001.
%%CITATION = HEP-TH 9803001;%%

\bibitem{Maldacena:1998im}
J.~Maldacena,
``Wilson Loops in Large $N$ Field Theories,''
Phys.\ Rev.\ Lett.\  {\bf 80}, 4859 (1998)
[hep-th/9803002].
%%CITATION = HEP-TH 9803002;%%

\bibitem{Berenstein:1999ij}
D.~Berenstein, R.~Corrado, W.~Fischler and J.~Maldacena,
``The Operator Product Expansion for Wilson Loops and Surfaces in the
Large $N$ Limit,''
Phys.\ Rev.\  {\bf D59}, 105023 (1999)
[hep-th/9809188].
%%CITATION = HEP-TH 9809188;%%

\bibitem{Drukker:1999zq}
N.~Drukker, D.~J.~Gross and H.~Ooguri,
``Wilson Loops and Minimal Surfaces,''
Phys.\ Rev.\  {\bf D60}, 125006 (1999)
[hep-th/9904191].
%%CITATION = HEP-TH 9904191;%%

\bibitem{Adler:1972qq}
S.~L.~Adler,
``Massless, Euclidean Quantum Electrodynamics on the Five-Dimensional
Unit Hypersphere,''
Phys.\ Rev.\  {\bf D6} (1972) 3445.
%%CITATION = PHRVA,D6,3445;%%

\bibitem{Drukker:2000ep}
N.~Drukker, D.~J.~Gross and A.~Tseytlin,
``Green-Schwarz String in $AdS_5\times S^5$: Semiclassical Partition
Function,''
JHEP {\bf 0004}, 021 (2000)
[hep-th/0001204].
%%CITATION = HEP-TH 0001204;%%

\bibitem{Mehta}
M.~L.~Mehta,
``A Method of Integration over Matrix Variables,''
Commun.\ Math.\ Phys.\  {\bf 79}, 327 (1981).
%%CITATION = CMPHA,79,327;%%

\bibitem{Kostov}
I.~K.~Kostov and M.~L.~Mehta,
``Random Surfaces of Arbitrary Genus: Exact Results for $D = 0$ And $-2$
Dimensions,''
Phys.\ Lett.\  {\bf B189}, 118 (1987).
%%CITATION = PHLTA,B189,118;%%

\bibitem{grotay}
D.~J.~Gross and W.~Taylor
``Twists and Wilson Loops in String Theory of Two Dimensional
$\, \rm  QCD_2$. ''
Nucl.\  Phys.\ {\bf B403}, 395 (1993).
%%CITATION = NUPHB,B403,395;%%

\bibitem{groog}
D.~J.~Gross and H.~Ooguri
``Aspects of Large $N$ Gauge Theory Dynamics as Seen by String Theory''
 Phys. Rev. {\bf D58},106002   (1998).
  %%CITATION =  PHRVA, D58,106002;%%

\end{thebibliography}
\end{document}